# SOXS: a wide band spectrograph to follow up transients

P. Schipani*[a], S. Campana[b], R. Claudi[c], H. U. Käufl[d], M. Accardo[d], M. Aliverti[b], A. Baruffolo[c], S. Ben-Ami[e], F. Biondi[c], A. Brucalassi[d,f], G. Capasso[a], R. Cosentino[g,h], F. D'Alessio[i], P. D'Avanzo[b], O. Hershko[j], D. Gardiol[k], H. Kuncarayakti[l,m], M. Munari[h], A. Rubin[j], S. Scuderi[h], F. Vitali[i], J. Achrén[n], J. Antonio Araiza-Duran[f], I. Arcavi[o], A. Bianco[b], E. Cappellaro[c], M. Colapietro[a], M. Della Valle[a], O. Diner[j], S. D'Orsi[a], D. Fantinel[c], J. Fynbo[p], A. Gal-Yam[j], M. Genoni[b], M. Hirvonen[q], J. Kotilainen[l,m], T. Kumar[m], M. Landoni[b], J. Lehti[q], G. Li Causi[r], D. Loreggia[k], L. Marafatto[c], S. Mattila[m], G. Pariani[b], G. Pignata[f], M. Rappaport[j], D. Ricci[c], M. Riva[b], B. Salasnich[c], R. Zanmar Sanchez[h], S. Smartt[s], M. Turatto[c]

[a]INAF - Osservatorio Astronomico di Capodimonte, Salita Moiariello 16, I-80131, Napoli, Italy
[b]INAF – Osservatorio Astronomico di Brera, Via Bianchi 46, I-23807 Merate (LC), Italy
[c]INAF – Osservatorio Astronomico di Padova, Vicolo dell'Osservatorio 5, I-35122 Padova, Italy
[d]ESO, Karl Schwarzschild Strasse 2, D-85748, Garching bei München, Germany
[e]Harvard Smithsonian Center for Astrophysics, Cambridge, USA
[f]Universidad Andres Bello, Avda. Republica 252, Santiago, Chile
[g]INAF - FGG, TNG, Rambla J.A. Fernández Pérez 7, E-38712 Breña Baja (TF), Spain
[h]INAF – Osservatorio Astronomico di Catania, Via S. Sofia 78 30, I-95123 Catania, Italy
[i]INAF – Osservatorio Astronomico di Roma, Via Frascati 33, I-00078 Monte Porzio Catone, Italy
[j]Weizmann Institute of Science, Herzl St 234, Rehovot, 7610001, Israel
[k]INAF – Osservatorio Astrofisico di Torino, Via Osservatorio 20, I-10025 Pino Torinese (TO), Italy
[l]Finnish Centre for Astronomy with ESO (FINCA), FI-20014 University of Turku, Finland
[m]Tuorla Observatory, Department of Physics and Astronomy, FI-20014 University of Turku, Finland
[n]Incident Angle Oy, Capsiankatu 4 A 29, FI-20320 Turku, Finland
[o]Tel Aviv University, Department of Astrophysics, 69978 Tel Aviv, Israel
[p]DARK Cosmology Center, Juliane Maries Vej 30, DK-2100 Copenhagen, Denmark
[q]ASRO (Aboa Space Research Oy), Tierankatu 4B, FI-20520 Turku, Finland
[r]INAF – Istituto di Astrofisica e Planetologia Spaziali, Via Fosso del Cavaliere, I-00133 Roma, Italy
[s]Astrophysics Research Centre, Queen's University Belfast, Belfast, County Antrim, BT7 1NN, UK

**ABSTRACT**

SOXS (Son Of X-Shooter) will be a spectrograph for the ESO NTT telescope capable to cover the optical and NIR bands, based on the heritage of the X-Shooter at the ESO-VLT. SOXS will be built and run by an international consortium, carrying out rapid and longer term Target of Opportunity requests on a variety of astronomical objects. SOXS will observe all kind of transient and variable sources from different surveys. These will be a mixture of fast alerts (e.g. gamma-ray bursts, gravitational waves, neutrino events), mid-term alerts (e.g. supernovae, X-ray transients), fixed time events (e.g. close-by passage of minor bodies). While the focus is on transients and variables, still there is a wide range of other astrophysical targets and science topics that will benefit from SOXS. The design foresees a spectrograph with a Resolution-Slit product $\approx$ 4500, capable of simultaneously observing over the entire band the complete spectral range from the U- to the H-band. The limiting magnitude of R~20 (1 hr at S/N~10) is suited to study transients identified from on-going imaging surveys. Light imaging capabilities in the optical band (grizy) are also envisaged to allow for multi-band photometry of the faintest transients. This paper outlines the status of the project, now in Final Design Phase.

*pietro.schipani@inaf.it

Table 1. SOXS Consortium Institutes

| Institute | Country |
|---|---|
| Istituto Nazionale di AstroFisica (INAF), PI institute | Italy |
| Department of Particle Physics and Astrophysics, Weizmann Institute of Science | Israel |
| University Andres Bello | Chile |
| FINCA - Finnish Centre for Astronomy with ESO & Turku University | Finland |
| Queen's University Belfast | UK |
| Tel Aviv University | Israel |
| Niels Bohr University | Denmark |

## 1. THE PROJECT

In 2015 ESO selected SOXS out of 19 proposals in response to the "Call for Scientific Projects for the ESO NTT on the La Silla Observatory", which had been issued in February 2014. SOXS has a key role in the new ESO strategy for the La Silla Observatory described in the ESO long term plan [1], that envisages the dedication of the two telescopes operated by ESO in La Silla to specific topics. They are the study of the transient sky with SOXS at the NTT and the radial velocity studies for exoplanets with HARPS and the new instrument NIRPS at the 3.6m telescope.

The research on transients has expanded significantly in the past decades, leading to some of the most recognized discoveries in astrophysics (e.g. gravitational wave events and their electromagnetic counterparts [2], gamma-ray bursts [3], super-luminous supernovae [4], accelerating universe [5][6]). As a sort of pathfinder to the SOXS science, a large fraction of the NTT observing time over the past few years has been dedicated to a public spectroscopic survey (the Public ESO Spectroscopic Survey of Transient Objects – PESSTO [7]) with 150 nights per year and a Large Programme (ePESSTO) with 200 nights over two years.

SOXS has very clear synergies with many other existing or upcoming major facilities. Many ground- or space-based facilities for searching new transients are operating, starting or are being planned in the near future. Among them are GAIA, PanSTARRS, Zwicky Transient Factory, LSST and EUCLID for optical searches, Swift, Fermi, SVOM, MAGIC and CTA for high-energy objects and, in the newly emerged field of non electromagnetic messengers, aLIGO/VIRGO for gravitational waves and KM3NET and ICECUBE for neutrinos. Nevertheless, so far most of the transient discoveries still lack an adequate spectroscopic follow-up. Thus, it is generally acknowledged that with the availability of so many transient factories in the next future, the scientific bottleneck is the spectroscopic follow-up observations of transients.

Within this context, SOXS aims to significantly contribute bridging this gap. It will be one of the few spectrographs on a dedicated telescope with a significant amount of observing time to characterize astrophysical transients. It is based on the concept of X-Shooter [8] at the VLT but, unlike its "father", the SOXS science case is heavily focused on transient events. Foremost, it will contribute to the classifications of transients, i.e. supernovae, electromagnetic counterparts of gravitational wave events, neutrino events, tidal disruptions of stars in the gravitational field of supermassive black holes, gamma-ray bursts and fast radio bursts, X-ray binaries and novae, magnetars, but also asteroids and comets, activity in young stellar objects, blazars and AGN.

SOXS will simultaneously cover the electromagnetic spectrum from 0.35 to 2.0 μm using two arms (UV-VIS and NIR) with a product slit-resolution of $\approx$ 4500. The throughput should enable to reach a S/N$\approx$10 in a 1-hour exposure of an R=20 mag point source. SOXS aims to see first light at the end of 2020.

The SOXS consortium is in charge of the realization of the instrument, with duties extending also over the next operation phase, within the framework of an agreement between ESO and SOXS. The consortium is expected to provide the user support through a helpdesk. This includes providing tools for the community (e.g. the ETC, tools to construct and submit the observing blocks). The consortium will further provide scheduling and merging of targets from the GTO and the regular ESO programmes. The transient programmes will be supported through dynamical scheduling and real-time interaction with the telescope operators. Data reduction to the 1D extracted spectrum will be done by the SOXS consortium and delivered to the ESO archive. The consortium will provide a public pipeline for the SOXS data.

In return of the efforts and investments, the SOXS consortium will be remunerated with 900 NTT nights over 5 years. The ESO community will access the rest of the NTT observing time. The consortium will be granted a proprietary period for their data. ESO will provide telescope operators and day-time maintenance and support on site.

The SOXS consortium structure has evolved since the proposal, including new partners. Now it is composed by the institutes listed in Table 1, located in six countries. The SOXS project went through a Preliminary Design Review in July 2017 and an anticipated Final Design Review for the detector systems and the optics in 2018. At the time of writing, the project is approaching the last Final Design Review step.

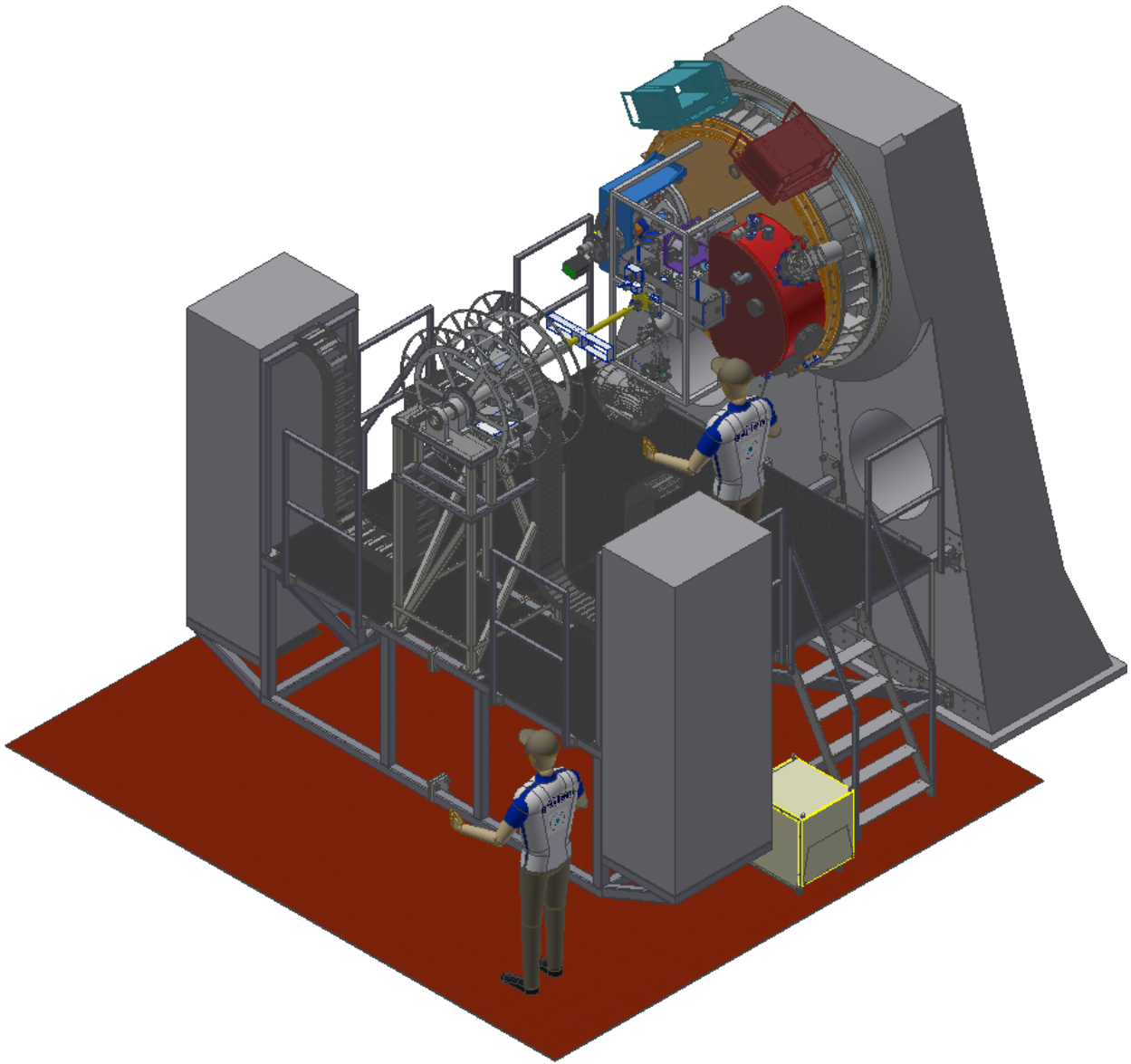

Figure 1. The SOXS instrument installed at the Nasmyth focus of the NTT.

## 2. THE INSTRUMENT

SOXS will be installed at one of the Nasmyth foci of the NTT (Figure 1, Figure 2), replacing SOFI.
Overall, the instrument is based on two distinct spectrographs, one operating in the UV-VIS 350-850 nm and the other in the NIR 800-2000 nm wavelength ranges. A small wavelength overlap has been designed to allow for an easier cross-calibration of the two spectrographs. The optical design has evolved since the first proposal [9]. The UV-VIS arm is based on a novel multi-grating concept, while the NIR part implements the 4C (Collimator Compensation of Camera Chromatism [10]) layout. The two arms are fed by the light coming through a common opto-mechanical system (the Common Path). It redirects the light from the telescope focus to the spectrograph slits through relay optics reducing the F/number and compensating for the atmospheric dispersion (only in the UV-VIS, where the dispersion at low altitude angles is unacceptably large). The Common Path provides also the mechanisms to drive the light to/from the other instrument subsystems, i.e. the acquisition camera and the calibration unit. Although the main operating mode for SOXS is the spectroscopy, the acquisition camera with a 3.5`×3.5` field allows also for a light imaging mode in the visible band. The SOXS calibration unit controls the lamps for the wavelength calibration and flat-fielding.

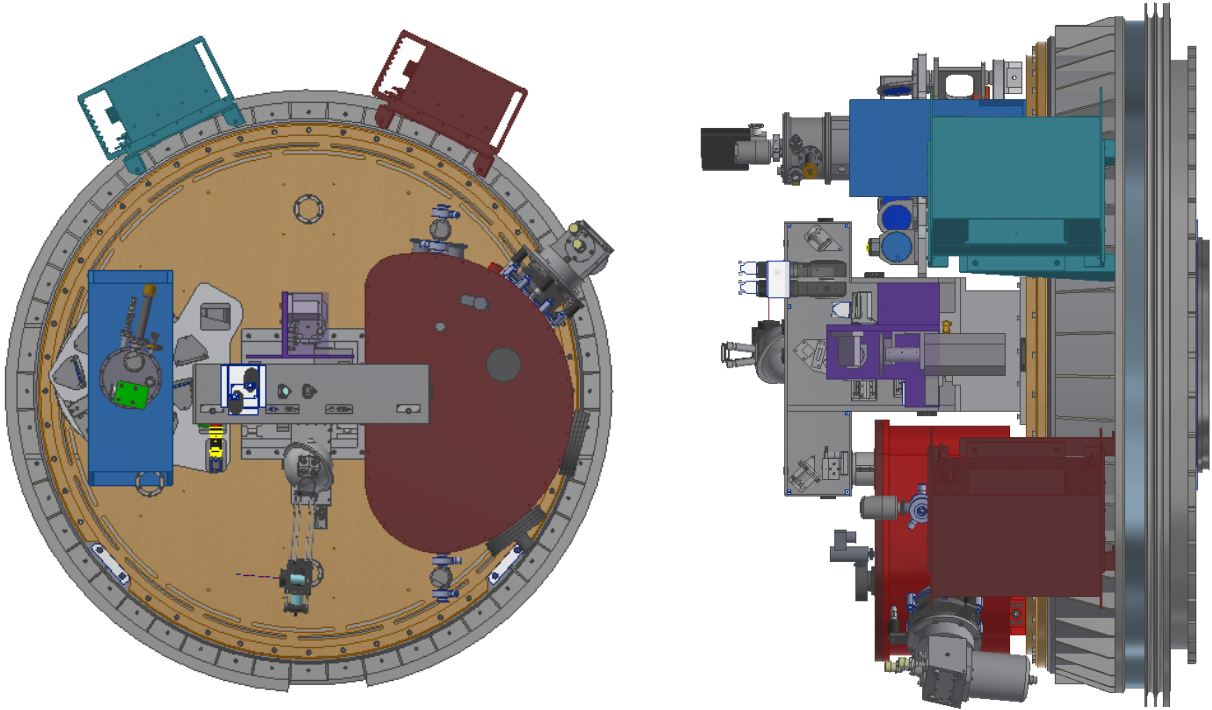

Figure 2. Front and side view of the instrument on the rotator flange.

The new instrument will be installed on the NTT adapter/rotator flange to compensate for the field rotation. A new dedicated Nasmyth platform was designed and will be installed at the telescope. Although an old platform was available, the new one will be smaller and tailored to the real instrument needs, allowing to host the control electronics in two symmetric racks and a co-rotator system to drive cables and pipes through two cable-wraps.

Figure 3 gives a functional description of the system. The light coming from the telescope goes through an instrument shutter used as a safety cover and for blocking any spurious light from the telescope side during calibrations. It is closed during tests and calibrations. Two stages allow for the operations of the calibration unit and the acquisition camera. The calibration stage allows to select between the light coming from the calibration and flat-field lamps, and the one from the telescope side. The calibration unit includes also a pinhole stage to create an artificial star.

The acquisition stage allows to select between several functionalities: sending the full field to the camera (acquisition and imaging mode); sending the light to the spectrographs through a 15″ hole and deviating peripheral light to the camera (spectroscopic mode); sending the calibration light to spectrographs through a pin-hole creating an artificial source, visible also on the camera (engineering mode); illuminating the slits with calibration light through a pellicle beam splitter allowing to see the slits on the camera (slit viewer mode).

The Common Path includes two folding mirrors equipped with tip-tilt devices to compensate for flexures, one per each spectrograph. The UV-VIS arm is equipped with an Atmospheric Dispersion Corrector implemented by two counter-rotating prism doublets, in order to minimize slit losses. The atmospheric dispersion in the NIR is considerably smaller (~0.5″ at 60° zenith angle) than in the UV-VIS; considering that the slit can be aligned along the dispersion direction at large zenith angles by offsetting the instrument rotator, we decided not to implement an ADC in the NIR.

Each spectrograph is equipped with a slit exchanger to accommodate slits of different widths (0.5″, 1″, 1.5″ for science; 5″ for calibration).

Most of the motorized functions are included in the Common Path. There is just one cryogenic motor (the NIR slit exchanger) in the instrument, in accordance to the guideline of reducing them to a minimum number. The telescope focus will be tuned on the UV-VIS arm. A warm focus mechanism in the NIR part of the Common Path will compensate for the focus difference between the two spectrographs, if any.

In order to simplify the maintenance of the instrument, the newest ESO standard solutions were preferred whenever possible. The SOXS control implementation for detectors, motors, vacuum and cryogenic systems will be easily integrated in the La Silla – Paranal Observatory and familiar to the technical staff.

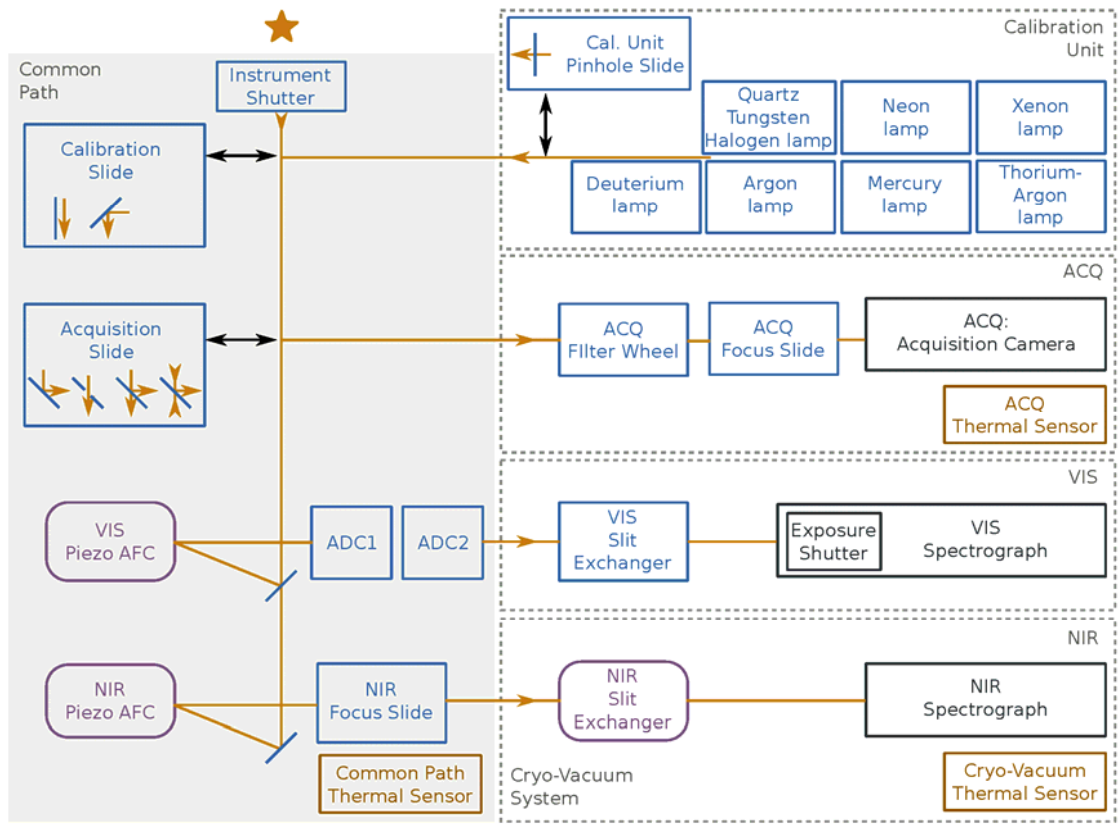

Figure 3. The SOXS functional diagram.

Table 2. Common Path parameters.

| Input F/N | F/11 |
|---|---|
| Field of view | 12x12 arcsec |
| Output F/N | F/6.5 |
| Image scale | 110 μm / arcsec |
| Wavelength range | 350-850 nm (UV-VIS); 800-2000 nm (NIR) |

**2.1 The Common Path**

The Common Path [11] relays the light from the NTT Focal Plane to the entrance of the two spectrographs (UV-VIS and NIR). It selects the wavelength range for the spectrographs using a dichroic and changes the focal ratio. A sketch of the common path opto-mechanical design is shown in in Figure 4 (overall dimensions are about 650x350 mm) and the main parameters in Table 2. After the dichroic, two flat folding mirrors direct the light towards two distinct UV-VIS and NIR arms. In the UV-VIS path, the light coming from the folding mirror goes to the ADC assembly, that strictly speaking is not composed only of two double prisms correcting the atmospheric dispersion but also of two doublets glued on the prisms. The two doublets (the first one having an aspherical surface) create a collimated beam for the ADC and transform the telescope F/11 beam into an F/6.5. After the ADC, the beam is reflected by the tip/tilt mirror mounted on a piezo stage. Finally, a field lens matches the exit pupil on the UV-VIS spectrograph pupil. The adopted glasses assure a good transmission (>80%) in the UV-VIS side of the spectrum. For the same reason, the dichroic is used in reflection for this wavelength range, in order to give the largest choice of materials for the substrate.

The near infrared path is very similar. It does not include an ADC as mentioned previously. A doublet reduces the telescope F/11 beam to an F/6.5 beam. The doublet has an aspherical surface deemed feasible by manufacturers. A flat tip-tilt folding mirror based on a piezo stage relays the light towards the slit. A flat window is used at the entrance of the spectrograph dewar, with a cold stop after the window itself to reduce the noise. A field lens, placed near the slit, remaps the telescope pupil on the grating of the spectrograph, as in the UV-VIS arm.

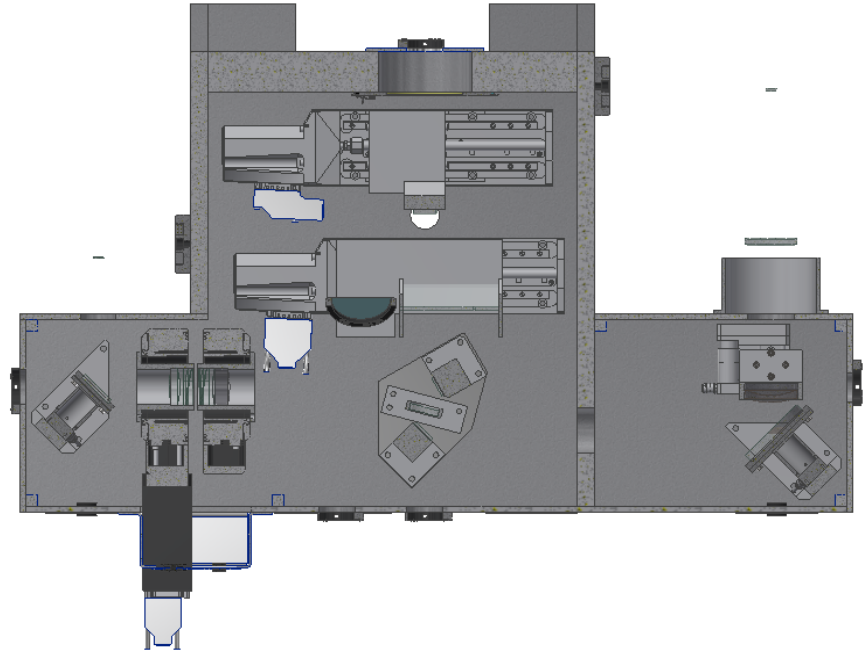

Figure 4. The Common Path optomechanics.

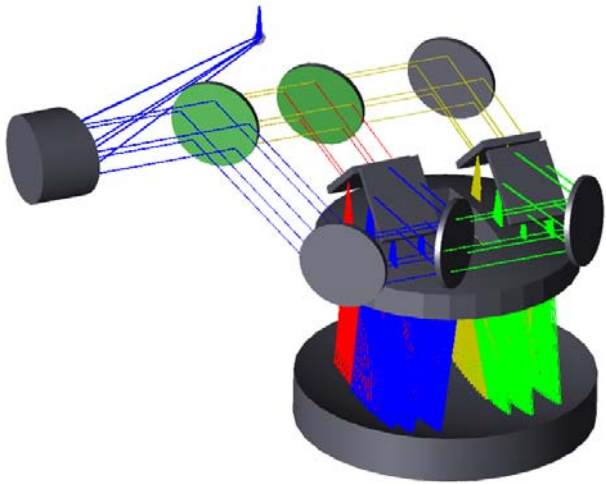

Figure 5. The UV-VIS spectrograph optical layout.

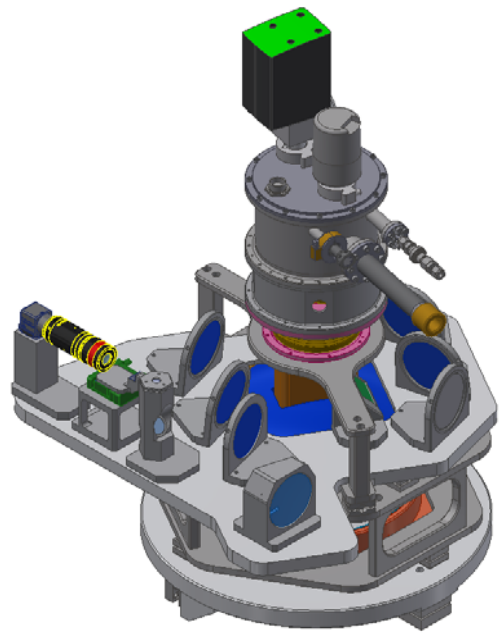

Figure 6. The UV-VIS optomechanics.

### 2.2 The UV-VIS spectrograph

The SOXS UV-VIS spectrograph [12] is based on a novel concept in which the incoming beam is partitioned into four polychromatic beams using dichroic surfaces, each covering a waveband range of ~100-200 nm. Each quasi-order is diffracted by an ion-etched grating. The four beams enter a three-element catadioptric camera that images them onto a common detector. The goal of the partitioning is to maximize the overall system throughput. The spectrograph is divided into two levels, both conceptually and mechanically (Figure 5, Figure 6).

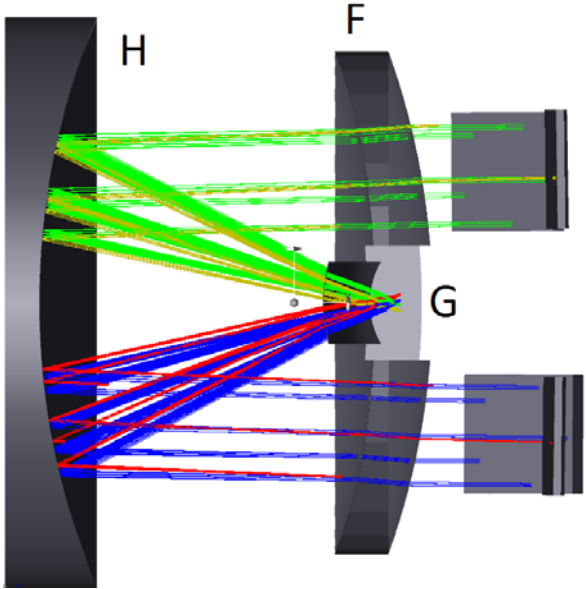 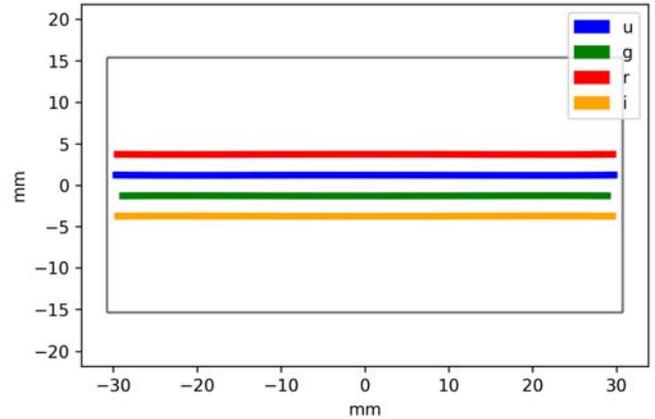

Figure 7. The UV-VIS spectrograph camera.   Figure 8. UV-VIS arm: spectral format.

Table 3. CCD characteristics.

| Detector | CCD44-82 |
|---|---|
| Chip type | Thinned back illuminated |
| Pixel size | 15 μm |
| Area | 2048 x 4096 pixels, 30.7 x 61.4 mm |
| QE at 500 nm | 90% |
| Coating | yes |
| Flatness | Better than 20 μm peak to valley |
| Peak signal | 200 ke$^-$/pixel |
| Charge transfer efficiency | 99.9995 % |
| Gain (e$^-$/ADU) | Low noise mode: 0.6 ± 0.1<br>Fast mode: 2 ± 0.2 |
| Detector Read-out Noise (RMS) | Low noise mode: <3 e$^-$<br>Fast mode: < 8e$^-$ |

The first level divides the beam into four collimated beams, each covering a different wavelength range. The beam from the Common Path is directed into the UV-VIS enclosure and folded by a flat pickup mirror located 30 mm downstream from the Common Path focus. The beam intercepts a ϕ=76.2 mm off-axis-parabolic mirror at an angle of 30°, creating a collimated beam with a diameter of ϕ=45 mm. The collimated beam is then partitioned into four beams using flat dichroic mirrors at ~45° angle of incidence.

The second level includes the dispersers and the camera that images the spectra of each quasi-order to a common detector. Each collimated beam intercepts its own ion-etched grating. The gratings are tilted by 45° w.r.t. the Nasmyth flange so that the incidence angle of the collimated beam is ~41°. The dispersed beams are then imaged by a catadioptric camera (Figure 7), made of three aspheric elements and inspired by the camera designed for MOONS [13], onto the common detector. The field flattener also serves as cryostat window, and is inserted through a rectangular aperture in the camera's corrector. The detector is located 4 mm behind the field flattener back surface. It is an e2V CCD44-82 CCD (Table 3), driven by the ESO NGC controller [14]. The CCD will be cooled adopting a Continuous Flow Cryostat.

The spectral format of the UV-VIS spectrograph (Figure 8) will be fairly regular. The main parameters are summarized in Table 5.

Table 4. NIR array characteristics.

| Detector | Teledyne H2RG |
|---|---|
| Detector Material | MBE HgCdTe double layer planar heterostructures (DLPH) |
| Substrate | CdZnTe substrate, removed to minimize fringing and optimize QE |
| Format | 2048x2048 |
| Pixel size | 18.0 μm |
| Number of outputs | 32 outputs for science frame access of internal bus outputs |
| Frame rate | 0.76 Hz using 32 parallel outputs |
| Reset by row | Non-destructive readout possible |
| Readout noise | Double correlated: < 20 e$^-$ rms<br>16 Fowler pairs < 7 e$^-$ rms |
| Storage capacity @ 0.5 V | 80 ke$^-$ |
| Dark current @ 77 K | < 0.1 e$^-$/s |

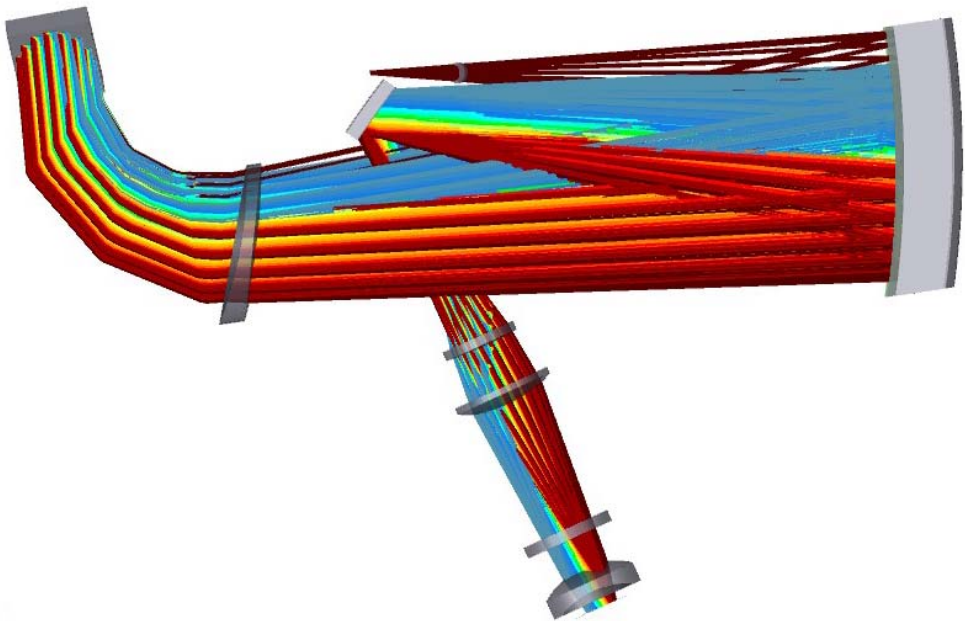

Figure 9. NIR spectrograph optical layout.

**2.3 The NIR Spectrograph**

The near infrared spectrograph [15] is a cross-dispersed echelle, with R=5000 (for 1 arcsec slit), covering the wavelength range from 800 to 2000 nm with 15 orders. It is based on the 4C concept, characterized by a very compact layout, reduced weight of optics and mechanics, good stiffness and high efficiency. The spectrograph is composed of a double pass collimator and a refractive camera, a grating disperser main and a prism-based cross disperser. The design is shown in Figure 9 and Figure 10, the main parameters are summarized in Table 5.

During the Final Design Phase the design evolved extending the wavelength range up to 2 μm, by adding a further order w.r.t. the PDR baseline. This decision caused few changes to the overall design, e.g. the working temperature of the spectrograph optomechanics was reduced from 180K to 150K in order to maintain adequate safety margins for the thermal radiation background, estimated to be well below the dark current of the array. The design includes an accurate baffling system as well as a (removable) thermal filter at 2 μm to cut off the longer wavelength radiation. The spectral format looks like a classical cross-dispersed echelle spectrum (Figure 11).

The detector is a Teledyne H2RG array (Table 4) operated at 40K, i.e. below its nominal working temperature of 77K, in order to take safe margins by design to avoid possible persistency problems. The vacuum and cryogenics system is based on a Closed Cycle Cryocooler, providing a more than adequate cooling power for our relatively small spectrograph volume.

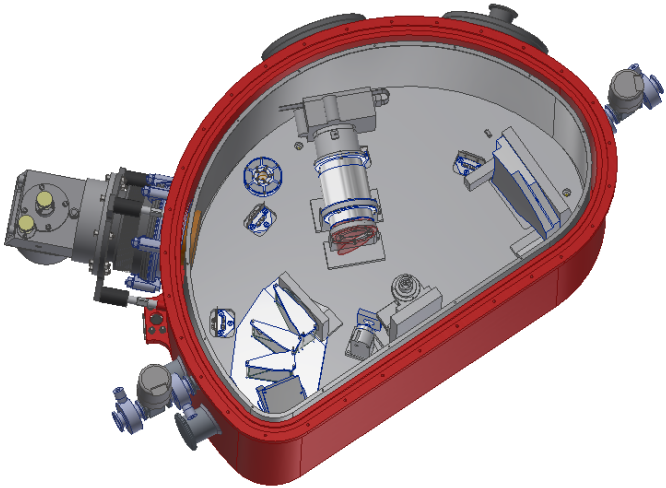
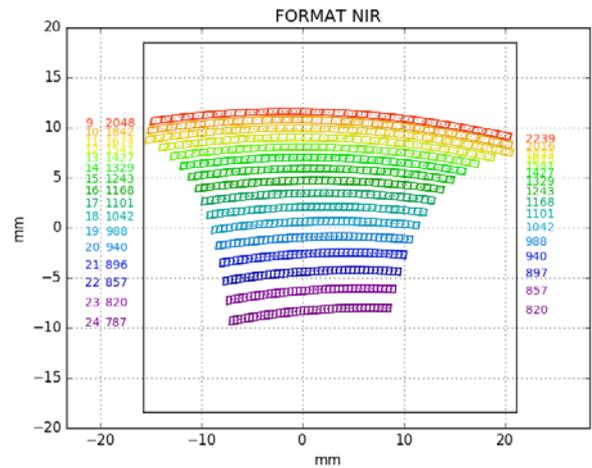

Figure 10. NIR optomechanics.

Figure 11. Near Infrared arm: echelle cross-dispersed spectral format.

Table 5. Main parameters of the spectrographs.

|  | UV-VIS | NIR |
|---|---|---|
| Collimator Focal Ratio | 6.5 | 6.5 |
| Collimator Beam diameter | 45 mm | 50 mm |
| Spectral range | 350-850 nm | 800-2000 nm |
| Resolution (1 arcsec slit) | 3500-7000 (≈4500 avg) | 5000 |
| Slit scale | 110 µm/arcsec | 110 µm/arcsec |
| Slit widths | 0.5 - 1 - 1.5 - 5 arcsec | 0.5 - 1 - 1.5 - 5 arcsec |
| Silt height | 12 arcsec | 12 arcsec |
| Camera Output Focal Ratio | 3.11 | 3.7 |
| Detector | e2V CCD44-82 2Kx4K | Teledyne H2RG 2Kx2K |
| Pixel Size | 15 µm | 18 µm |
| Detector Scale | 0.28 arcsec/pixel | 0.25 arcsec/pixel |
| Main Disperser | Four custom ion etched gratings | Grating 44° blaze angle, 4° off-plane |
| Cross Disperser | N/A | 3 Cleartran Prisms, apex angle 20°, 20°, 26° |
| Working temperature | Ambient (-5C, +20C) | 150K (40K for the last two elements of the camera and detector) |

## 2.4 Acquisition Camera

The Acquisition Camera [16] shown in Figure 12 will primarily work for the acquisition of the target, allowing for the centering of the objects on the slits. Additionally, it will provide imaging capability for the visible band in a 3.5x3.5 arc minute field, and the engineering monitoring of the co-alignment between the spectrographs. The main components of the system are a CCD detector, the focal reducer optics, a re-focuser linear stage, a flat 45° folding mirror configurable in different positions (spectroscopy mode, acquisition or imaging mode, artificial star mode), an 8 position filter wheel with 6 LSST (u, g, r, i, z, y) + V Johnson filters and a free position for the slit viewer. The F/number of the camera is 3.6, the pixel scale is 0.205 arcsec/pixel. All the system is included in an aluminum structure.

The camera will further provide a secondary guiding capability. The telescope will guide using its own guiding system but will be able to accept secondary guiding corrections from SOXS. The telescope control software has built-in commands to handle this functionality, coordinating the two guiding sources.

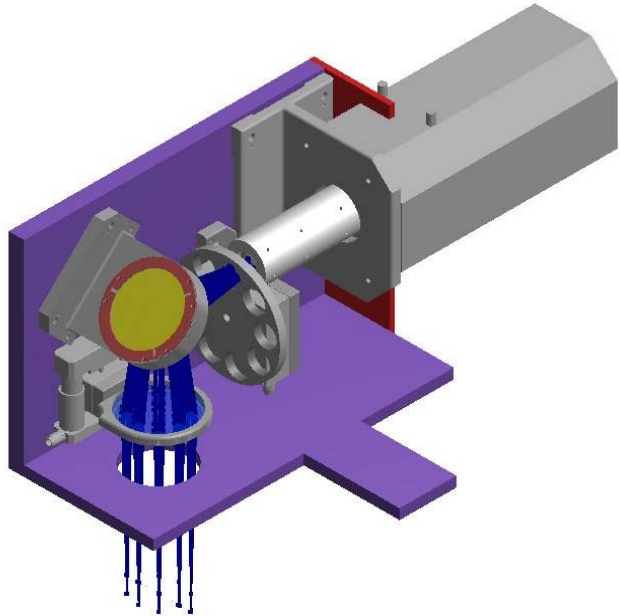

Figure 12. Acquisition Camera layout.

The camera detector will be the Andor iKon M934, using a CCD sensor BEX2-DD (Back Illuminated CCD, Deep Depletion with fringe suppression, extended range dual AR coating) 1024x1024 with 13μm pixel size, low readout noise, negligible dark current with industry-leading thermoelectric cooling down to -100°C.

### 2.5 Calibration Unit

The calibration unit (Figure 13) is designed to provide the calibration spectra to remove the instrument signatures and convert the observed spectrum into one with physical units (in wavelength/frequency and flux). The calibration spectra are generated using a synthetic light source, adopting an integrating sphere equipped with lamps suitable for wavelength and flux calibrations across the full wavelength range of the instrument (350-2000 nm). The following lamps are used:
- Quartz-tungsten-halogen (QTH) lamp, for flux calibration 500-2000 nm
- Deuterium (D2) lamp, for flux calibration 350-500 nm (used simultaneously with QTH lamp for UV-VIS arm flux calibration)
- NeArHgXe penray lamps bundled together, for NIR wavelength calibration. The individual lamps are controlled to operate together as one lamp
- ThAr hollow cathode lamp, for UV-VIS wavelength calibration

The diffuse light exiting the integrating sphere is directed by the relay optics to achieve uniform illumination of the slit, while mimicking a beam of light exiting from the telescope. In addition, a fold mirror is foreseen in the light path to accommodate the optical design to the space and mechanical constraints. A stage-controlled calibration pick-up mirror is placed in the light path before the telescope focus, which enables the observer to switch the spectrograph input light between telescope and calibration sources. The system is also equipped with a pinhole mask mounted on a linear motor stage, located near the exit port of the integrating sphere. This enables the creation of an artificial star for engineering purposes. The lamp enclosure is equipped with a safety interlock system, which cuts off electricity when the cover is opened. This system is intended to prevent UV radiation from the lamps from harming the service personnel.

### 2.6 Vacuum & Cryogenics

The SOXS Vacuum system (Figure 14) is designed with pre-vacuum and turbo-molecular pumps that are common to the two arms. They will be located on the Nasmyth platform at a very short distance and connected only when vacuum operations are needed. This solution reduces the load on the instrument, leaving the possibility to connect the turbo-molecular pump on board if needed. The commercial components have been selected in agreement to ESO standards. In the visible arm, the small CCD cryostat will be based on the Continuous Flow Cryostat concept, successfully adopted in several ESO projects. A Closed Cycle CryoCooler will be adopted in the near infrared instrument cryogenic system.

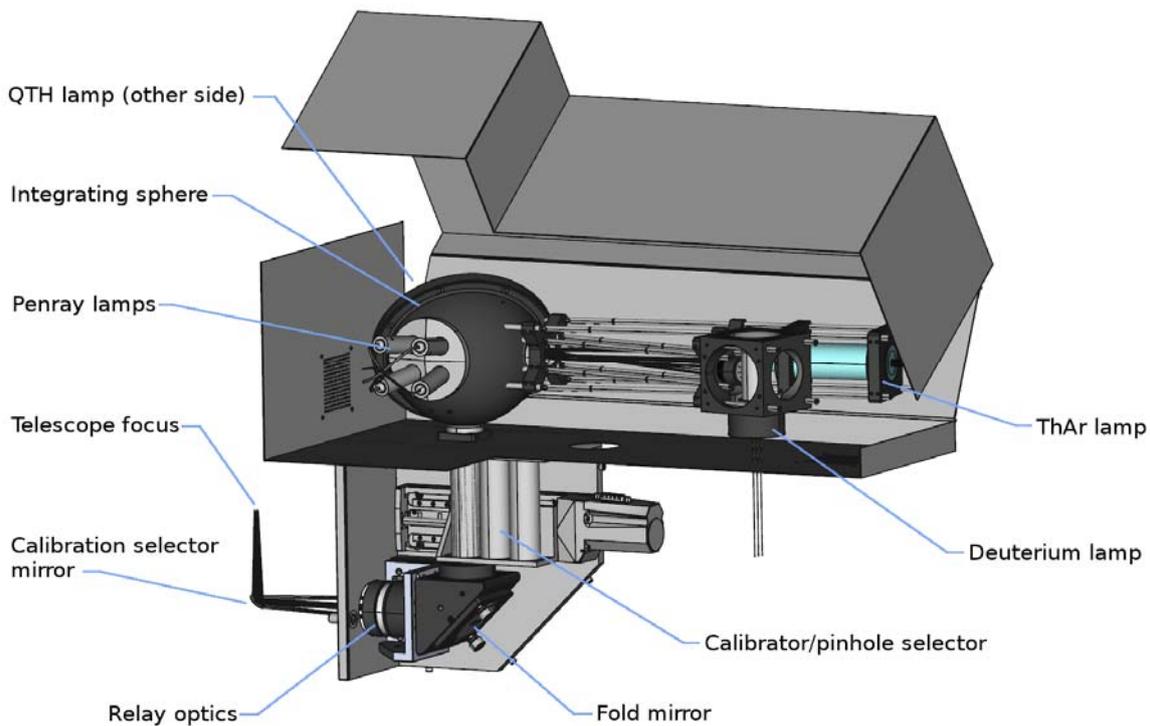
Figure 13. Calibration Unit.

**2.7 Control System**

In the SOXS project, the consortium had total freedom in the choice of control system software and hardware technologies. Nevertheless, the previous experiences in projects for ESO and the perspective to build something to install in an ESO Observatory led to adopt the same VLT / ELT standards, both for hardware and software.

The software [17] is based on the VLT Instrumentation Common Software. This solution led to standard management of instrument components, observation, calibration and maintenance procedures. Observation procedures for spectroscopy, imaging, calibration and maintenance are developed as templates, which are executed by the *Broker of Observation Blocks* (BOB) through commands sent to the Observation Software. The work is in progress for the usual software subsystems, i.e. the Instrument Control Software, the Detector Control Software, and the Observation and Maintenance Software.

The electronics [18] consists of the following subsystems.

- Instrument electronics, based on Beckhoff PLC technology, according to the most recent developments for ESO instrumentation. The control of all motors and actuators is included in this category.
- Detector System electronics, based on the ESO New Generation Controllers. It controls the two detectors and the shutter in the visible arm.
- Vacuum and Cryogenic system electronics, based on Siemens PLC technology, according to the ESO standards. This part of electronics is designed to work independently of the instrument software, that communicates only for monitoring.
- Calibration Unit Electronics. It is based on Beckhoff PLC as well, but it is physically a separate subsystem because it is developed in a separate institute and this solution allows to manage better the interface.

## 3. CONCLUSIONS

SOXS will be the work-horse for the study of transients of the ESO La Silla observatory. The project is going to conclude the final design phase in 2018. Afterward, the plan is to have the first light within 2020. After the completion of the instrument, the consortium will still be involved in the regular operation phase. Further details on specific subsystems can be found in [11]-[12] and [14]-[21].

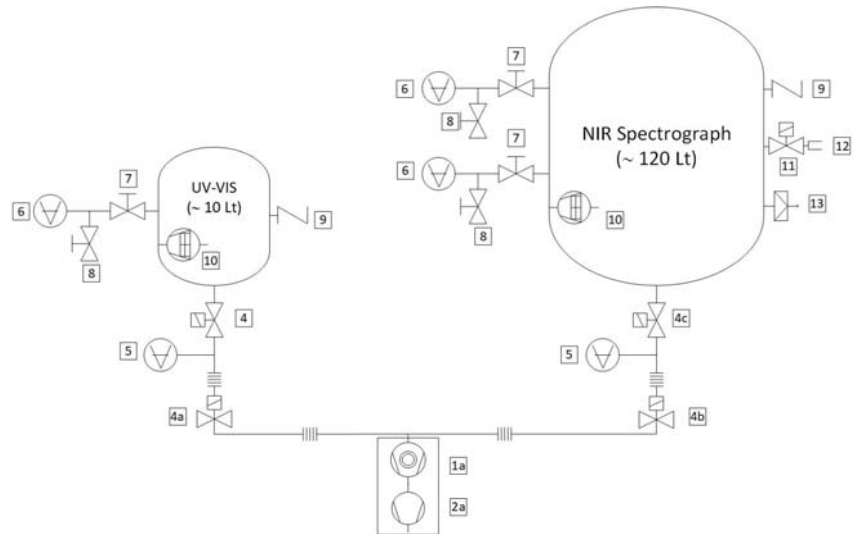

Figure 14. Vacuum system.